\input ./style/arxiv-general.cfg
\documentclass[aap,MSNbibl,seceqn,dvips]{arximspdf}
\makeatletter
   \@ifpackageloaded{graphicx}{}{\usepackage{graphicx}}
\makeatother
\usepackage{mathrsfs}

\doi{10.1214/15-AAP1103}
\volume{26}
\issue{2}
\pubyear{2016}
\firstpage{794}
\lastpage{817}
\docsubty{FLA}

\makeatletter
\newcommand{\rrvert}{\vert}
\newcommand{\llvert}{\vert}
\def\implies{\Longrightarrow}
\newcommand{\eqref}[1]{(\ref{#1})}
\newtheorem{Theorem}{Theorem}[section] %
\newtheorem{Lemma}[Theorem]{Lemma} %
\newtheorem{Proposition}[Theorem]{Proposition} %

\newproclaim{Assumption}[Theorem]{Assumption}%
\newproclaim{Definition}[Theorem]{Definition}%
\newproclaim{Example}[Theorem]{Example} %
\newproclaim{Remark}[Theorem]{Remark} %

\newtheorem{Theoremm}{Theorem}[section] %
\newproclaim{Remarkk}[Theoremm]{Remark} %
\newtheorem{Lemmaa}[Theoremm]{Lemma}

\newcommand{\set}{\triangleq}
\newcommand{\bmo}{\mathrm{BMO}}
\newcommand{\Hbmo}{\mathcal{H}_{\bmo}} %
\newcommand{\Hp}{\mathcal{H}_{p}} %
\newcommand{\Hinfty}{\mathcal{H}_{\infty}} %

\newcommand{\sign}{\operatorname{sign}}
\newcommand{\trace}{\operatorname{trace}}
\newcommand{\const}{\operatorname{const}}

\renewcommand{\mathcal}{\mathscr}

\renewcommand{\epsilon}{\varepsilon}

\newcommand{\esssup}{\mathop{\operatorname{ess\,sup}}}

\makeatother

\begin{document}
\begin{frontmatter}

\title{A system of quadratic BSDEs arising in a price impact model}
\runtitle{BSDE in a price impact model}

\begin{aug}
\author[A]{\fnms{Dmitry} \snm{Kramkov}\thanksref{t1}\ead[label=e1]{kramkov@cmu.edu}}
\and
\author[B]{\fnms{Sergio} \snm{Pulido}\corref{}\ead[label=e2]{sergio.pulidonino@ensiie.fr}}
\runauthor{D. Kramkov and S. Pulido}
\affiliation{Carnegie Mellon University and
ENSIIE \& Universit\'e d'\'Evry-Val-d'Essonne, LaMME, UMR CNRS 8071}
\thankstext{t1}{The author also holds a part-time
position at the University of Oxford. Supported
in part by the Carnegie Mellon-Portugal Program and by the
Oxford-Man Institute for Quantitative Finance at the University of Oxford.}
\address[A]{Department of Mathematical Sciences\\
Carnegie Mellon University\\
5000 Forbes Avenue\\
Pittsburgh, Pennsylvania 15213-3890\\
USA\\
\printead{e1}}
\address[B]{Laboratoire de Math\'ematiques\\
et Mod\'elisation d'\'Evry (LaMME)\\
23 Boulevard de France\\
91037 \'Evry Cedex\\
France\\
\printead{e2}}
\end{aug}

%
\received{\smonth{8} \syear{2014}}
%
\revised{\smonth{1} \syear{2015}}

%
\begin{abstract}
We consider a financial model where the prices of risky assets are
quoted by a representative market maker who takes into account an
exogenous demand. We characterize these prices in terms of a system
of BSDEs with quadratic growth. We show that this system admits a
unique solution for every bounded demand if and only if the market
maker's risk-aversion is sufficiently small. The uniqueness is
established in the natural class of solutions, without any
additional norm restrictions. To the best of our knowledge, this is
the first study that proves such (global) uniqueness result for a
system of fully coupled quadratic BSDEs.
\end{abstract}

%
\begin{keyword}[class=AMS]
\kwd{60H10}
\kwd{91B24}
\kwd{91G80}
\end{keyword}
%
%
\begin{keyword}
\kwd{Liquidity}
\kwd{price impact}
\kwd{multi-dimensional quadratic BSDE}
\end{keyword}
\end{frontmatter}

\section{Introduction}
\label{sec:1}

In the classical problem of optimal investment, an economic agent
trades at exogenous stock prices and looks for a strategy maximizing
his expected utility. This problem has been extensively studied in
the literature with various approaches. For example, Merton \cite{Mert:71}
relied on PDEs, Kramkov and Schachermayer \cite{KramSch:03} used the
methods of convex duality
and martingales and Hu et al. \cite{HuImkelMull:05} employed BSDEs.

In this paper, we consider an inverse problem: find stock prices for
which a given strategy is optimal; that is, instead of the usual task
of getting ``(optimal stocks') quantities from prices'' we want to
deduce ``prices from quantities.'' This problem naturally arises in
the market microstructure theory; see Grossman and Miller \cite{GrosMill:88},
Garleanu et al.
\cite{GarlPederPotes:09} and German \cite{Germ:11}. Here, the strategy
represents the continuous demand on the market for a set of
divided-paying stocks. The representative dealer, with exponential
utility, provides liquidity for these assets and quotes prices in such
a way that the market clears. In \cite{GarlPederPotes:09} and
\cite{Germ:11}, the existence and uniqueness of such prices is
established for every \emph{simple} demand process, where trades occur
only a finite number of times. It is the purpose of this paper to
cover the general case.

As a first step, we obtain in Theorem~\ref{th:1} an equivalent
characterization of the demand-based prices in terms of solutions to a
system of BSDEs with quadratic growth. Similar systems appear
naturally in economic equilibrium problems with exponential
preferences; see Frei and dos Reis \cite{FreiReis:11}. Contrary to the
one-dimensional
case, which is well studied and where general criteria for existence
and uniqueness are available (see, e.g., Kobylanski \cite{Kobyl:00} and
Briand and Hu \cite{BrianHu:06}), the situation with a \emph{system}
of quadratic
BSDEs is more delicate. A counter-example in \cite{FreiReis:11} shows
that, in general, such system may not have solutions even for a
\emph{bounded} terminal condition. Moreover, although the existence
can be guaranteed when the values at maturity are sufficiently small
(see Proposition~1 in Tevzadze \cite{Tevz:08}), the uniqueness is only obtained
in a local manner.

Our main results are stated in Theorem~\ref{th:2} and
Proposition~\ref{prop:1}. In Theorem~\ref{th:2}, we prove that the
solutions to our system of quadratic BSDEs exist and are (globally)
unique, provided that the product of the $\bmo$-norm of the stocks'
dividends, the $\mathbf{L}_\infty$-norm of the demand and the
dealer's risk-aversion is sufficiently small. To the best of our
knowledge, this is the first study that proves a (global) uniqueness
result for a system of fully coupled quadratic BSDEs. In
Proposition~\ref{prop:1}, we show that, in general, such well-posedness
may be violated even if the dividends and the demand are bounded. A
crucial role in our study is played by the ``sharp'' {a priori}
estimate given in Lemma~\ref{lem:2}. This estimate is obtained
considering the stochastic control problem, which corresponds to the
maximization of the dealer's expected utility with respect to demands
bounded by $1$.

\subsection*{Notation}
\label{sec:notations-1}

For a matrix $A = (A^{ij})$, we denote its transpose by $A^*$ and
define its norm as
\[
\llvert A\rrvert \set\sqrt{\trace{AA^*}} = \sqrt{\sum
_{i,j} \bigl(A^{ij} \bigr)^2}.
\]

We will work on a filtered probability space $(\Omega, \mathcal{F},
(\mathcal{F}_t)_{t\in[0,T]}, \mathbb{P})$ satisfying the standard
conditions of right-continuity and completeness; the initial
$\sigma$-algebra $\mathcal{F}_0$ is trivial,
$\mathcal{F}=\mathcal{F}_T$, and the maturity $T$ is finite. The
expectation is denoted as $\mathbb{E}[\cdot]$ and the conditional
expectation with respect to $\mathcal{F}_t$ as $\mathbb{E}_t[\cdot]$.

We shall use the following spaces of stochastic processes:
\begin{longlist}[$\bmo(\mathbf{R}^m)$]
\item[$\bmo(\mathbf{R}^m)$] is the Banach space of continuous
$m$-dimensional martingales $M$ with $M_0=0$ and the norm
\[
\Vert M\Vert_{\bmo} \set\sup_{\tau} \bigl\Vert \bigl\lbrace{
\mathbb{E}_\tau \bigl[\llvert M_T-M_\tau\rrvert
^2 \bigr]} \bigr\rbrace^{1/2}\bigr\Vert_{\infty},
\]
where the supremum is taken with respect to all stopping times
$\tau$.
\item[$\mathcal{H}_0(\mathbf{R}^{m\times d})$] is the vector space of
predictable processes $\zeta$ with values in $m\times d$-matrices
such that $\int_0^T \llvert \zeta_s\rrvert ^2 \,ds < \infty$.
This is precisely
the space of $m\times d$-dimensional integrands $\zeta$ for a
$d$-dimensional Brownian motion $B$. We shall identify $\alpha$ and
$\beta$ in $\mathcal{H}_0(\mathbf{R}^{m\times d})$ if $\int_0^T
\llvert \alpha_s- \beta_s\rrvert ^2 \,ds = 0$ or,
equivalently, if the
stochastic integrals $\alpha\cdot B$ and $\beta\cdot B$ coincide.
\item[$\Hp(\mathbf{R}^{m\times d})$] for $p\geq1$ consists of
$\zeta
\in\mathcal{H}_0(\mathbf{R}^{m\times d})$ such that
\[
\Vert\zeta\Vert_p \set \biggl\lbrace{\mathbb{E} \biggl[ \biggl(\int
_0^T \llvert \zeta _s\rrvert
^2\,ds \biggr)^{p/2} \biggr]} \biggr\rbrace^{1/p} <
\infty.
\]
It is a complete Banach space under this norm.
\item[$\Hbmo(\mathbf{R}^{m\times d})$] consists of $\zeta\in
\mathcal{H}_0(\mathbf{R}^{m\times d})$ such that $\zeta\cdot B \in
\bmo(\mathbf{R}^m)$ for a $d$-dimensional Brownian motion $B$. It is
a Banach space under the norm:
\[
\Vert\zeta\Vert_{\bmo} \set\Vert\zeta\cdot B\Vert_{\bmo} = \sup
_{\tau} \biggl\Vert \biggl\lbrace{\mathbb{E}_\tau \biggl[
\int_\tau^T \llvert \zeta_s\rrvert
^2\,ds \biggr]} \biggr\rbrace^{1/2}\biggr\Vert_{\infty}.
\]
\item[$\mathcal{H}_\infty(\mathbf{R}^{n})$] is the Banach space of
bounded $n$-dimensional predictable processes $\gamma$ with the
norm:
\[
\Vert\gamma\Vert_{\infty} \set\inf \bigl\lbrace{c\geq 0}\dvtx \bigl\llvert
\gamma_t(\omega) \bigr\rrvert \leq c, dt\times \mathbb {P}[d\omega]
\mbox{-a.s.} \bigr\rbrace.
\]
\end{longlist}

For an $n$-dimensional integrable random variable $\xi$ with
$\mathbb{E}[\xi] = 0$ set
%
\begin{equation}
\label{eq:1} \Vert\xi\Vert_{\bmo} \set\bigl\Vert \bigl(\mathbb{E}_t[
\xi] \bigr)_{t\in
[0,T]}\bigr\Vert_{\bmo} .
\end{equation}
Denote also
\begin{eqnarray*}
\Vert\xi\Vert_p &\set& \bigl(\mathbb{E} \bigl[\llvert \xi\rrvert
^p \bigr] \bigr)^{1/p},\qquad p\geq1,
\\
\Vert\xi\Vert_\infty&\set&\inf \bigl\lbrace{c\geq0}\dvtx \bigl\llvert \xi(
\omega) \bigr\rrvert \leq c, \mathbb{P}[d\omega]\mbox{-a.s.} \bigr\rbrace.
\end{eqnarray*}
Observe that
%
\begin{equation}
\label{eq:2} \Vert\xi\Vert_{\bmo} \leq\inf_{x\in\mathbf{R}^n}
\Vert\xi -x\Vert_\infty.
\end{equation}

\section{Model}
\label{sec:2}

There is a single \emph{representative} market maker whose preferences
regarding terminal wealth are modeled by the exponential utility with
the risk aversion coefficient $a>0$:
%
\begin{equation}
\label{eq:3} U(x) = -\frac{1}a e^{-ax},\qquad x\in\mathbf{R}.
\end{equation}
The financial market consists of a bank account and $n$ stocks. The
bank account pays an \emph{exogenous} interest rate, which we assume
to be zero. The stocks pay dividends $\Psi= (\Psi^i)_{i=1,\ldots,n}$
at maturity $T$; each $\Psi^i$ is a random variable. While the
terminal stocks' prices $S_T$ are always given by $\Psi$, their values
$S_t$ on $[0,T)$ are determined \emph{endogenously} by the equilibrium
mechanism specified below; in particular, they are affected by demand
on stocks. Following Garleanu et al. \cite{GarlPederPotes:09} and
German \cite{Germ:11}, we
give the following definition.

\begin{Definition}
\label{def:1}
A predictable process $\gamma= (\gamma_t)$ with values in
$\mathbf{R}^n$ is called a \emph{demand}. The demand $\gamma$ is
\emph{viable} if there is an $n$-dimensional semimartingale of
\emph{stock prices} $S=(S_t)$ such that $S_T=\Psi$, the probability
measure $\mathbb{Q}$, called the \emph{pricing measure}, is
well defined by
\[
\frac{d\mathbb{Q}}{d\mathbb{P}} \set\frac{U'(\int_0^T \gamma
\,dS)}{\mathbb{E}[U'(\int_0^T \gamma
\,dS)]} = \frac{e^{-a\int_0^T \gamma
\,dS}}{\mathbb{E}[e^{-a\int_0^T \gamma \,dS}]},
\]
and $S$ and the stochastic integral $\gamma\cdot S$ are uniformly
integrable martingales under~$\mathbb{Q}$.
\end{Definition}

In this definition, $-\gamma_t$ stands for the number of stocks that
an external counter-party plans to buy/sell from the market up to time
$t$. The stochastic integral $\gamma\cdot S$ represents the evolution
of the losses of the external counter-party or, equivalently, of the
gains of the market maker. Note that, as $S=S(\gamma)$, the
dependence of $\gamma\cdot S$ on $\gamma$ is \emph{nonlinear}; this is
in contrast to the standard, ``small agent's,'' model of mathematical
finance.

To clarify the economic meaning of Definition~\ref{def:1}, we recall a
well-known result in the theory of optimal investment, which states
that under the stock prices $S=S(\gamma)$ the strategy $\gamma$ is
optimal.

\begin{Lemma}
\label{lem:1}
Let the utility function $U$ be given by \eqref{eq:3} and $\gamma$
be a viable demand accompanied by the stock prices $S$ and the
pricing measure $\mathbb{Q}$ in the sense of
Definition~\ref{def:1}. Then
\[
\mathbb{E} \biggl[U \biggl(\int_0^T\gamma \,dS
\biggr) \biggr] \geq \mathbb{E} \biggl[U \biggl(\int_0^T
\zeta \,dS \biggr) \biggr],
\]
for every demand $\zeta$ such that the stochastic integral $\zeta
\cdot S$ is a $\mathbb{Q}$-supermartingale.
\end{Lemma}

\begin{pf}
Define the conjugate function to $U$:
\[
V(y)\set\sup_{x\in\mathbf{R}} \bigl\{U(x)-xy \bigr\} = \frac{1}a
y( \ln y - 1), \qquad y>0,
\]
and observe that, as
\[
V \bigl(U'(x) \bigr) = U(x) - xU'(x),\qquad x\in
\mathbf{R},
\]
the construction of $\mathbb{Q}$ yields that
%
\begin{equation}
\label{eq:4} V \biggl(y\frac{d\mathbb{Q}}{d\mathbb{P}} \biggr) = U \biggl(\int
_0^T \gamma \,dS \biggr) - y\frac{d\mathbb{Q}}{d\mathbb{P}}
\int_0^T \gamma \,dS,
\end{equation}
where
\[
y = \mathbb{E} \biggl[U' \biggl(\int_0^T
\gamma \,dS \biggr) \biggr] = \mathbb{E} \bigl[e^{-a \int_0^T \gamma \,dS} \bigr].
\]
On the other side, clearly,
%
\begin{equation}
\label{eq:5} V \biggl(y\frac{d\mathbb{Q}}{d\mathbb{P}} \biggr) \geq U \biggl(\int
_0^T \zeta \,dS \biggr) - y\frac{d\mathbb{Q}}{d\mathbb{P}} \int
_0^T \zeta \,dS.
\end{equation}
Taking expectations (under $\mathbb{P}$) in \eqref{eq:4}
and \eqref{eq:5}, we obtain the conclusion.
\end{pf}

We call a demand $\gamma$ \emph{simple} if
\[
\gamma=\sum_{i=0}^{m-1}
\theta_{i}1_{(\tau_i,\tau_{i+1}]},
\]
where $0=\tau_0 < \tau_1 < \cdots< \tau_m = T$ are stopping times and
$\theta_i$ is a $\mathcal{F}_{\tau_i}$-measu\-rable random variable with
values in $\mathbf{R}^n$, $i=0,\ldots,m-1$. Theorem~1 in
\cite{Germ:11} shows that if the dividends $\Psi= (\Psi^i)$ have all
exponential moments, then every bounded simple demand $\gamma$ is
viable. Moreover, the price process $S=S(\gamma)$ is unique and is
constructed explicitly, by backward induction.

The goal of this paper is to investigate the case of demands $\gamma$
with general continuous dynamics. Our main results,
Theorem~\ref{th:2} and Proposition~\ref{prop:1}, rely on the
BSDE-characterization of the stock prices $S=S(\gamma)$ from the next
section.

\begin{Remark}
\label{rem:1}
To simplify notation, we neglected in our setup the existence of
the initial random endowment $\beta_0$ for the market maker. Due to
the choice of exponential utility in \eqref{eq:3}, this condition
does not restrict any generality. Indeed, if $\beta_0 \neq0$, then,
in Definition~\ref{def:1} and throughout the paper, the measure
$\mathbb{P}$ should just be replaced by the measure $\mathbb{Q}(0)$
with the density
\[
\frac{d\mathbb{Q}(0)}{d\mathbb{P}} \set \frac{U'(\beta_0)}{\mathbb{E}[U'(\beta_0)]} = \frac{\exp(-a\beta_0)}{\mathbb{E}[\exp(-a\beta_0)]}.
\]
\end{Remark}

\section{Characterization in terms of BSDE}
\label{sec:char-terms-bsde}

Hereafter, we shall assume that:
\begin{longlist}[(A1)]
\item[(A1)]
There exists a $d$-dimensional Brownian motion
$B$ such that every local martingale $M$ is a stochastic integral
with respect to $B$:
\[
M = M_0 + \zeta\cdot B.
\]
\end{longlist}
Of course, this assumption holds if the filtration is generated by
$B$.

For a viable demand $\gamma$ accompanied by stocks' prices $S$ define
the process $R$ such that
%
\begin{equation}
\label{eq:6} R_t \set U^{-1} \biggl(
\mathbb{E}_t \biggl[U \biggl(\int_t^T
\gamma \,dS \biggr) \biggr] \biggr)
\\
= - \frac{1}a \log \bigl(\mathbb{E}_t \bigl[e^{-a \int_t^T \gamma
\,dS}
\bigr] \bigr),
\end{equation}
is the market maker's \emph{certainty equivalent value} at time $t$ of
the \emph{remaining gain} $\int_t^T \gamma \,dS$. Observe that the
density process $Z$ of the pricing measure $\mathbb{Q}$ has the form
\[
Z_t \set\mathbb{E}_t \biggl[ \frac{d\mathbb{Q}}{d\mathbb{P}} \biggr] =
e^{-a(R_t - R_0 +
\int_0^t \gamma \,dS)},\qquad t\in[0,T].
\]
Jensen's inequality and the martingale property of $\gamma\cdot S$
under $\mathbb{Q}$ imply that $Z^{-1}e^{-aR} = e^{-a (R_0 -
\gamma\cdot S)}$ is a $\mathbb{Q}$-submartingale. Hence, $e^{-aR}$
is a submartingale (under $\mathbb{P}$) and, as $R_T=0$, we obtain
that
%
\begin{equation}
\label{eq:7} e^{-aR} \leq1\quad \mbox{or, equivalently}\quad R\geq0.
\end{equation}

Under (A1), there is $\alpha\in\mathcal{H}_0(\mathbf{R}^d)$,
the \emph{market price of risk}, such that
\[
Z = \mathcal{E}(- \alpha\cdot B) = e^{-\alpha\cdot B -({1}/2) \int
\llvert \alpha\rrvert ^2 \,dt}.
\]
From Girsanov's theorem, we deduce that
\[
W \set B + \int\alpha \,dt
\]
is a Brownian motion under $\mathbb{Q}$ and that every local
martingale under $\mathbb{Q}$ is a stochastic integral with respect to
$W$. In particular, there is $\sigma\in
\mathcal{H}_0(\mathbf{R}^{n\times d})$, the \emph{volatility} of
stocks' prices, such that
\[
S = S_0 + \sigma\cdot W = S_0 + \int\sigma\alpha \,dt +
\sigma\cdot B.
\]

We now characterize $S$, $R$, $\alpha$ and $\sigma$ in terms of
solutions to the multi-dimensional quadratic
BSDE \eqref{eq:8}--\eqref{eq:9}.

\begin{Theorem}
\label{th:1}
Assume \textup{(A1)}. An $n$-dimensional predictable process $\gamma$
is a viable demand if and only if there are processes
$(S,R,\eta,\theta)$, where $S$ is a $n$-dimensional semi-martingale,
$R$ is a semi-martingale, $\eta\in\mathcal{H}_0(\mathbf{R}^d)$, and
$\theta\in\mathcal{H}_0(\mathbf{R}^{n\times d})$, such that, for
every $ t\in[0,T]$,
%
\begin{eqnarray}
\label{eq:8} aR_t &=& \frac{1}2 \int_t^T
\bigl( \bigl\llvert \theta^*_s\gamma_s \bigr\rrvert
^2 - \llvert \eta_s\rrvert ^2 \bigr) \,ds -
\int_t^T \eta \,dB,
\\
\label{eq:9} aS_t &=& a\Psi- \int_t^T
\theta_s \bigl(\eta_s+\theta_s^*
\gamma_s \bigr) \,ds - \int_t^T
\theta \,dB,
\end{eqnarray}
and such that the stochastic exponential $Z \set\mathcal{E}(-(\eta
+ \theta^*\gamma)\cdot B)$ and the processes $ZS$ and $Z(\gamma\cdot
S)$ are (uniformly integrable) martingales.

In this case, $S$ represents stocks' prices which accompany
$\gamma$, $R$ is the certainty equivalent value, $Z$ is the density
process of the pricing measure $\mathbb{Q}$, and the market price of
risk $\alpha$ and the volatility $\sigma$ are given by
%
\begin{eqnarray}
\label{eq:10} \alpha&=& \eta+ \theta^*\gamma,
\\
\label{eq:11} \sigma&=& \theta/a.
\end{eqnarray}
\end{Theorem}

\begin{pf}
Let $\gamma$ be a viable demand accompanied by stocks' prices $S$
and the certainty equivalent value $R$. Define the martingales
\begin{eqnarray*}
L_t &=& \mathbb{E}_t \biggl[U' \biggl(\int
_0^T \gamma \,dS \biggr) \biggr] =
\mathbb{E}_t \bigl[e^{-a\int_0^T
\gamma \,dS} \bigr],
\\
M_t &=& a\mathbb{E}_t \biggl[\Psi U'
\biggl(\int_0^T \gamma \,dS \biggr) \biggr] = a
\mathbb{E}_t \bigl[\Psi e^{- a\int_0^T
\gamma \,dS} \bigr],
\end{eqnarray*}
and observe that the pricing measure $\mathbb{Q}$ has the density
$L_T/L_0$ and
\begin{eqnarray*}
aS_t &=& a\mathbb{E}_t^{\mathbb{Q}}[\Psi] =
M_t/L_t,
\\
aR_t &=& aR_0 - \log(L_t/L_0) -
\int_0^t \gamma d(M/L),
\end{eqnarray*}
or, in a ``backward'' form, as $S_T = \Psi$ and $R_T = 0$,
\begin{eqnarray*}
aS_t &=& a\Psi- \int_t^T \,d(M/L),
\\
aR_t &=& \int_t^T \bigl(d\log L +
\gamma d(M/L) \bigr).
\end{eqnarray*}
From (A1) and accounting for the strict positivity of $L$, we
deduce the existence and uniqueness of $\alpha\in
\mathcal{H}_0(\mathbf{R}^d)$ and $\beta\in
\mathcal{H}_0(\mathbf{R}^{n\times d})$ such that
\begin{eqnarray*}
L &=& L_0 - L\alpha\cdot B,
\\
M &=& M_0 + L\beta\cdot B.
\end{eqnarray*}
Direct computations based on  It\^o's formula yield
\begin{eqnarray*}
d\log L &= &-\frac{1}2 \llvert \alpha\rrvert ^2 \,dt - \alpha
dB,
\\
d(M/L) &=& \biggl(\beta\alpha+ \frac{1}L M\llvert \alpha\rrvert
^2 \biggr)\,dt + \biggl(\beta+\frac{1}L M\alpha^* \biggr)\,dB
\\
&=& \bigl(\beta+ aS\alpha^* \bigr)\alpha \,dt + \bigl(\beta+aS\alpha^* \bigr)\,dB,
\end{eqnarray*}
and equations \eqref{eq:8} and \eqref{eq:9} hold with
\begin{eqnarray*}
\theta&=& \beta+ aS\alpha^*,
\\
\eta&=& \alpha- \theta^*\gamma.
\end{eqnarray*}

Observe that
\[
Z = \mathcal{E} \bigl(- \bigl(\eta+ \theta^*\gamma \bigr)\cdot B \bigr) =
\mathcal{E}(-\alpha\cdot B) = L/L_0
\]
is the density process of $\mathbb{Q}$ and, in particular, is a
martingale. The martingale properties of $ZS$ and $Z(\gamma\cdot S)$
under $\mathbb{P}$ then follow from those of $S$ and $\gamma\cdot S$
under $\mathbb{Q}$. Hence, the process $(S,R,\theta,\eta)$ satisfies
the conditions of the theorem.

Conversely, let $(S,R,\theta,\eta)$ be as in the statement of the
theorem. Define the probability measure $\mathbb{Q}$ with the
density process $Z= \mathcal{E}(-(\eta+ \theta^*\gamma)\cdot
B)$. From~\eqref{eq:8} and \eqref{eq:9}, we deduce that
\begin{eqnarray*}
\frac{d\mathbb{Q}}{d\mathbb{P}} & =& Z_T = e^{-\int_0^T(\eta+
\theta^*\gamma)\,dB - ({1}/2) \int_0^T\llvert \eta+ \theta
^*\gamma\rrvert ^2\,dt}
\\
&=& e^{-a(R_T - R_0 + \int_0^T \gamma \,dS)} = \frac{U' (\int_0^T
\gamma \,dS )}{\mathbb{E} [U' (\int_0^T \gamma
\,dS ) ]}.
\end{eqnarray*}
Moreover, $S_T = \Psi$ and the martingale properties of $S$ and
$\gamma\cdot S$ under $\mathbb{Q}$ follow from those of $ZS$ and
$Z(\gamma\cdot S)$ under $\mathbb{P}$. Hence, $S$ satisfies the
conditions of Definition~\ref{def:1}.

Finally, as part of the arguments above, we obtained that, given the
stocks' prices $S$, the linear invertibility relations \eqref{eq:10}
and \eqref{eq:11} between $(\eta,\theta)$ and $(\alpha,\sigma)$ hold
and equations \eqref{eq:6} and \eqref{eq:8} for $R$ are
equivalent.
\end{pf}

\begin{Remark}
\label{rem:2}
The BSDE characterization in Theorem~\ref{th:1} heavily relies on
condition \eqref{eq:3} of exponential preferences. For a general
utility function $U$, one can similarly associate with the stock
prices $S=S(\gamma)$ the following system of Forward--Backward
Stochastic Differential Equations (FBSDEs):
\begin{eqnarray*}
S_t &=& \Psi-\int_t^T
\sigma_u\alpha_u \,du-\int_t^T
\sigma \,dB,
\\
Y_t &= &\log \bigl(U'(X_T) \bigr) +
\frac{1}{2}\int_t^T\llvert
\alpha_u\rrvert ^2 \,du+ \int_t^T
\alpha \,dB,
\\
X_t &=& \int_0^t\gamma \,dS.
\end{eqnarray*}
Here, $X$ is the gain process of the market maker due to the demand
$\gamma$ and $Y = \log Z + \const$ is a normalized log-density
process of the pricing measure. If $U$ is of exponential type then,
by ``decoupling'' substitution \eqref{eq:6}, this fully coupled
system of FBSDEs can be reduced to the simpler
system \eqref{eq:8}--\eqref{eq:9} of quadratic BSDEs.
\end{Remark}

\section{Existence and uniqueness}
\label{sec:main-results}

This is our main result.

\begin{Theorem}
\label{th:2}
Assume \textup{(A1)}. There is a constant $c=c(n)>0$ (dependent only
on the number of stocks $n$) such that if $\gamma\in
\Hinfty(\mathbf{R}^{n})$ and
%
\begin{equation}
\label{eq:12} a \Vert\gamma\Vert_\infty\bigl\Vert\Psi-\mathbb{E}[\Psi]
\bigr\Vert_{\bmo
} \leq c,
\end{equation}
then $\gamma$ is a viable demand accompanied by the unique stocks'
prices $S$. Moreover, the $\bmo$-norms of the volatility $\sigma$
and of the market price of risk $\alpha$ are bounded by
%
\begin{eqnarray}
\label{eq:13} \Vert\sigma\Vert_{\bmo} &\leq&2\bigl \Vert\Psi-\mathbb{E}[\Psi]
\bigr\Vert _{\bmo} ,
\nonumber
\\[-8pt]
\\[-8pt]
\nonumber
\Vert\alpha\Vert_{\bmo} &\leq&4 a \Vert\gamma\Vert_\infty\bigl \Vert
\Psi-\mathbb{E}[\Psi]\bigr\Vert_{\bmo} .
\end{eqnarray}
\end{Theorem}

As the following simple example illustrates, among the dividends
$\Psi$ with finite $\bmo$-norm, condition \eqref{eq:12} is
necessary even for the viability of constant demands.

\begin{Example}
\label{ex:1}
Suppose that $\Psi$ is a real-valued random variable such that
\[
\mathbb{E}[\Psi]=0, \qquad\Vert\Psi\Vert_{\bmo} <\infty\qquad \mbox{but } \mathbb{E}
\bigl[e^\Psi \bigr] =\infty;
\]
see, for example, Example~3.4 in Kazamaki \cite{Kazam:94}. It readily
follows from
Definition~\ref{def:1} that the constant demand $\gamma=-1/a$ is not
viable. Indeed, in this case, the pricing measure $\mathbb{Q}$ can
only be of the form:
\[
\frac{d\mathbb{Q}}{d\mathbb{P}} = \const e^{\Psi},
\]
which is not possible, because of the lack of integrability.
\end{Example}

It is more delicate to construct a counter-example for \emph{bounded}
dividends $\Psi$. Let $c=c(n)>0$ be a constant from
Theorem~\ref{th:1}. In view of \eqref{eq:2},
condition \eqref{eq:12} holds if
\[
a \Vert\gamma\Vert_\infty\inf_{x\in\mathbf{R}^n}\Vert\Psi -x
\Vert_\infty \leq c.
\]
The following proposition shows that, already in one-dimensional case,
the assertions of Theorem~\ref{th:2} may fail for {bounded} $\Psi$ and
that $c(1)<1$. It is stated under a stronger assumption
than (A1):
\begin{longlist}[(A2)]
\item[(A2)]
There exists a one-dimensional Brownian motion $B$
such that the filtration $(\mathcal{F}_t)$ is the completion of the
filtration generated by $B$:
\[
\mathcal{F}_t = \mathcal{F}_t^B \vee
\mathcal{N}^{\mathbb{P}},\qquad t\in[0,T].
\]
Here, $\mathcal{F}_t^B \set\sigma \lbrace{B_s,s\leq t}
\rbrace$ and
$\mathcal{N}^{\mathbb{P}}$ is the family of all $\mathbb{P}$-null
sets in $\mathcal{F}$. 
\end{longlist}

\begin{Proposition}
\label{prop:1}
Assume \textup{(A2)}. There exist a bounded predictable process
$\gamma$ and a bounded random variable $\Psi$ (both $\gamma$ and
$\Psi$ have dimension one) such that
\[
a \Vert\gamma\Vert_\infty\Vert\Psi\Vert_\infty\leq1,
\]
and such that $\gamma$ is not supported by a unique semi-martingale
$S$ in the sense of Definition~\ref{def:1}.
\end{Proposition}

Note that, in comparison to the nonexistence construction in
Example~\ref{ex:1} for dividends with finite $\bmo$-norm, our result
for bounded dividends is weaker. Here we only claim either
nonexistence or nonuniqueness.

\begin{Remark}
\label{rem:3}
In the follow-up paper \cite{KramkovPulido:14b}, we show that
under \eqref{eq:12} the prices $S=S(\gamma)$ are stable under small
changes in the demand $\gamma$; in particular, they can be well
approximated by the prices originated from simple demands. We also
obtain in \cite{KramkovPulido:14b} a power series expansion of
$S=S(\gamma,a)$ with respect to the market's risk-aversion $a$ in a
neighborhood of the point $a=0$ where the price impact effect
disappears.
\end{Remark}

\subsection{Outline of the proof of Theorem \texorpdfstring{\protect\ref{th:2}}{4.1}}
\label{sec:sketch-proof-theorem}

For the reader's convenience, we begin with an outline of the key
steps in the proof of Theorem~\ref{th:2}. To simplify notation,
suppose that
\[
\mathbb{E}[\Psi] = 0,\qquad a=1\quad \mbox{and}\quad \llvert \gamma\rrvert \leq1.
\]
By Theorem~\ref{th:2}, the existence and uniqueness of the price
process $S$, which accompanies the demand $\gamma$, is equivalent to
the existence and uniqueness of the solution $(\eta,\theta)$ of the
multi-dimensional quadratic BSDE:
\begin{eqnarray*}
R_t &=& \frac{1}2 \int_t^T
\bigl( \bigl\llvert \theta^*_s\gamma_s \bigr\rrvert
^2 - \llvert \eta_s\rrvert ^2 \bigr) \,ds -
\int_t^T \eta \,dB,
\\
S_t &=& \Psi- \int_t^T
\theta_s \bigl(\eta_s + \theta^*_s
\gamma_s \bigr) \,ds - \int_t^T
\theta \,dB,
\end{eqnarray*}
such that the stochastic exponential $Z \set\mathcal{E}(-(\eta+
\theta^*\gamma)\cdot B)$ and the processes $ZS$ and $Z(\gamma\cdot S)$
are martingales.

The first step is standard. Using a rather straightforward extension
of the results of Tevzadze \cite{Tevz:08} (see Theorem~\ref{th:3} in
the \hyperref[sec:bsde-with-quadratic]{Appendix}), we deduce the existence of a
constant $b=b(n)$ such that if
\[
\Vert\Psi\Vert_{\bmo} \leq b,
\]
then the BSDE admits only one solution $(\eta,\theta)$ such that
\[
\bigl\Vert(\eta,\theta)\bigr\Vert_{\bmo} \leq2b.
\]
Local existence and \emph{local} uniqueness then readily follow.

The delicate part is to verify the \emph{global} uniqueness. For that,
we need to find a constant $0<c\leq b$ such that
\[
\Vert\Psi\Vert_{\bmo} \leq c\quad \implies\quad\bigl\Vert(\eta,\theta)\bigr\Vert
_{\bmo} \leq2b,
\]
for \emph{every} solution $(\eta,\theta)$ for which
$Z=\mathcal{E}(-\int(\eta+ \theta^*\gamma)\,d B)$, $ZS$, and $Z
(\gamma
\cdot S)$ are martingales. Using basic $\bmo$-inequalities, we first
deduce the existence of an increasing function $f = f(x)$, $x\geq0$,
such that
\[
\bigl\Vert(\eta,\theta)\bigr\Vert_{\bmo} \leq f\bigl(\Vert R\Vert_{\infty} \bigr)
\Vert\Psi\Vert_{\bmo} .
\]
To conclude the argument, we need to find a constant $K>0$ and an
increasing function $g= g(x)$ on $[0,K)$, such that
\[
\Vert R\Vert_{\infty} \leq g\bigl(\Vert\Psi\Vert_{\bmo} \bigr) \qquad\mbox{if }
\Vert\Psi\Vert_{\bmo} < K.
\]

A sharp version of the above {a priori} estimate is obtained in
Lemma~\ref{lem:2} and is based on the verification arguments for the
stochastic control problem:
\[
u^*_t \set\esssup_{\llvert \gamma\rrvert \leq1} \bigl(-e^{-R_t(\gamma)} \bigr) =
\esssup_{\llvert \gamma\rrvert \leq1} \mathbb {E}_t \bigl[-e^{-\int_t^T
\gamma \,dS(\gamma)} \bigr],
\]
where we maximize the market maker's expected utility over all viable
demands $\gamma$ with $\llvert \gamma\rrvert \leq1$. Later,
this estimate is
also used in Proposition~\ref{prop:1} to produce a counter-example.

\subsection{Proof of Theorem \texorpdfstring{\protect\ref{th:2}}{4.1}}
\label{sec:proof-theor-refth:2}

From Definition~\ref{def:1}, we deduce that the dependence of stocks'
prices $S = S(\gamma,a,\Psi)$ on the viable demand $\gamma$, on the
risk-aversion coefficient $a$, and on the dividend $\Psi$ has the
following homogeneity properties: for $b>0$,
%
\begin{equation}
\label{eq:14} S(b\gamma,a,\Psi) = S(\gamma,ba,\Psi) = \frac{1}b S(
\gamma,a,b\Psi).
\end{equation}
This yields similar properties of the volatilities $\sigma=
\sigma(\gamma,a,\Psi)$ and of the market prices of risk $\alpha=
\alpha(\gamma,a,\Psi)$ which correspond to $S=S(\gamma,a,\Psi)$:
%
\begin{eqnarray}
\label{eq:15} \sigma(b\gamma,a,\Psi) &=& \sigma(\gamma,ba,\Psi) =
\frac{1}b \sigma(\gamma,a,b\Psi),
\nonumber
\\[-8pt]
\\[-8pt]
\nonumber
\alpha(b\gamma,a,\Psi) &=& \alpha(\gamma,ba,\Psi) = \alpha(\gamma,a,b\Psi).
\end{eqnarray}
In view of these identities, it is sufficient to prove
Theorem~\ref{th:2} for the case
%
\begin{equation}
\label{eq:16} a = 1 \geq\Vert\gamma\Vert_\infty.
\end{equation}

Define the function $H=H(u)$ on $[0,\infty)$ as
\[
H(u) = e^u (u-1) + 1,\qquad u\geq0.
\]
Observe that $H$ is an $N$-function in the theory of Orlicz spaces,
that is, it is convex, strictly increasing, $H(0) = H'(0)=0$, and
$H'(\infty)=\infty$; see Krasnosel'ski{\u\i} and Ruticki{\u\i}
\cite{KrasRutic:61}. For a later use, we
also note that for any $\epsilon>0$ there is a constant
$C(\epsilon)>0$ such that
%
\begin{equation}
\label{eq:17} \frac{1}2 u^2 \leq H(u) \leq C(\epsilon)
e^{(1+\epsilon)u},\qquad u\geq0.
\end{equation}

For an $n$-dimensional martingale $M$ with $M_0=0$ set
\[
\Vert M\Vert_{H} \set \inf \biggl\lbrace{\lambda>0}\dvtx \sup
_{\tau}\biggl\Vert\mathbb{E}_\tau \biggl[H \biggl(
\frac{ \llvert  M_T-M_\tau\rrvert }{\lambda
} \biggr) \biggr]\biggr\Vert_{\infty} \leq1 \biggr\rbrace,
\]
where the upper bound is taken with respect to all stopping times
$\tau$. Observe that, by the monotone convergence theorem,
%
\begin{equation}
\label{eq:18} \sup_{\tau}\biggl \Vert\mathbb{E}_\tau
\biggl[ H \biggl(\frac{\llvert  M_T-M_\tau
\rrvert }{\Vert M\Vert_{H}} \biggr) \biggr]\biggr\Vert_{\infty} \leq1.
\end{equation}
The family of $n$-dimensional martingales $M$ with $M_0=0$ and
$\Vert M\Vert_H<\infty$ is a Banach space under $\Vert\cdot\Vert
_{H}$ and
this norm is equivalent to the $\bmo$-norm: there is a constant $C_H=
C_H(n)>0$ such that
%
\begin{equation}
\label{eq:19} \tfrac{1}{\sqrt{2}} \Vert M\Vert_{\bmo} \leq\Vert M
\Vert_{H} \leq C_H \Vert M\Vert_{\bmo} .
\end{equation}
Here, the first inequality follows from the left-hand side
of \eqref{eq:17}, while the second one holds by Remark~2.1 on page 28
of Kazamaki \cite{Kazam:94}.

For an $n$-dimensional integrable random variable $\xi$ with
$\mathbb{E}[\xi] = 0$ denote
\[
\Vert\xi\Vert_{H} \set\bigl\Vert \bigl(\mathbb{E}_t[\xi]
\bigr)_{t\in[0,T]}\bigr\Vert_{H}.
\]

\begin{Lemma}
\label{lem:2}
Let $\gamma\in\mathcal{H}_\infty(\mathbf{R}^n)$ be a viable demand
accompanied by stocks' prices $S$ and the certainty equivalent value
$R$. Assume \textup{(A1)}, \eqref{eq:16} and that
\[
\mathbb{E}[\Psi] = 0,\qquad \Vert\Psi\Vert_{H} < 1.
\]
Then for every $x\in\mathbf{R}^n$ the process
\[
V_t(x) \set \bigl(1 - H \bigl(\llvert S_t-x\rrvert
\bigr) \bigr) e^{-R_t},\qquad t\in [0,T],
\]
is a supermartingale and the following estimate holds:
%
\begin{equation}
\label{eq:20} e^{-R_t} \geq1 - \Vert\Psi\Vert_{H},\qquad t
\in[0,T].
\end{equation}
\end{Lemma}

\begin{pf}
To simplify notation, set
\[
F(u) \set1 - H(u) = e^{u}(1 - u),\qquad u\geq0.
\]
As the density process of the pricing measure $\mathbb{Q}$ has the
form:
\[
Z_t \set \mathbb{E}_t \biggl[\frac{d\mathbb{Q}}{d\mathbb{P}} \biggr] =
e^{-(R_t -
R_0 + \int_0^t \gamma \,dS)},\qquad t\in[0,T],
\]
the $\mathbb{P}$-supermartingale property of $V(x)$ is equivalent to
the $\mathbb{Q}$-supermartin\-gale property of
\[
\widetilde{V}(x) \set e^{R_0}Z^{-1} V(x) = F \bigl(\llvert
S-x \rrvert \bigr) e^{\gamma
\cdot
S}.
\]

Recall that under $\mathbb{Q}$ the price process $S$ evolves as
\[
S = S_0 + \sigma\cdot W,
\]
where $W$ is a Brownian motion under $\mathbb{Q}$. Using the fact
that $F'(0)=0$, we deduce from  It\^o's formula that
\[
\widetilde{V}_t(x) = M_t(x) + \int_0^t
e^{(\gamma\cdot
S)_r}A_r(x) \,dr,
\]
where $M(x)$ is a local martingale under $\mathbb{Q}$ and
\begin{eqnarray*}
A(x) &=& 1_{ \lbrace{\llvert  S-x\rrvert >0} \rbrace
} \biggl(\frac{1}2 F''
\bigl(\llvert S-x\rrvert \bigr) \frac{{\llvert \sigma^* (S-x)\rrvert ^2}}{\llvert
S-x\rrvert ^2} + \frac{1}2 F \bigl(
\llvert S-x\rrvert \bigr) \bigl\llvert \sigma^*\gamma \bigr\rrvert ^2
\\
&&\hspace*{50pt}{}+ F' \bigl(\llvert S-x\rrvert \bigr) \biggl(\frac{ \langle\sigma
^*(S-x),\sigma^*\gamma  \rangle}{\llvert  S-x\rrvert } \\
&&\hspace*{120pt}{}+
\frac{1}{2\llvert  S-x\rrvert } \biggl(\llvert \sigma \rrvert ^2 -
\frac{\llvert \sigma^*(S-x)\rrvert ^2}{\llvert  S-x\rrvert ^2} \biggr) \biggr) \biggr).
\end{eqnarray*}
As $\Vert\gamma\Vert_\infty\leq1$, $F'\leq0$, and
\[
F - 2 {F'} + F'' = 0,
\]
we deduce that
\[
A(x) \leq1_{ \lbrace{\llvert  S-x\rrvert >0}
\rbrace} \frac{\llvert \sigma\rrvert ^2}2 \bigl(F''
- 2{F'} + F \bigr) \bigl(\llvert S-x\rrvert \bigr) = 0,
\]
thus proving the local supermartingale property of
$\widetilde{V}(x)$ under $\mathbb{Q}$.

To verify that $\widetilde{V}(x)$ is a (global)
$\mathbb{Q}$-supermartingale, it is sufficient to show that
$\widetilde{V}(x)$ is bounded below by some
$\mathbb{Q}$-martingale. With this goal in mind, take $\epsilon>0$
such that
\[
\Vert\Psi\Vert_H < \frac{1}{1+\epsilon} < 1
\]
and observe that, by the construction of the norm $\Vert\cdot\Vert_H$,
\[
\mathbb{E} \bigl[e^{(1+\epsilon)\llvert \Psi\rrvert } \bigr] < \infty.
\]
It follows that
\[
\mathbb{E}^{\mathbb{Q}} \bigl[e^{(1+\epsilon)\llvert \Psi\rrvert  +
(\gamma\cdot S)_T} \bigr] = e^{R_0}
\mathbb{E} \bigl[e^{(1+\epsilon)\llvert \Psi\rrvert } \bigr] < \infty
\]
and hence, the $\mathbb{Q}$-martingale
\[
N_t \set\mathbb{E}^{\mathbb{Q}}_t
\bigl[e^{(1+\epsilon)\llvert \Psi
\rrvert  +
(\gamma\cdot S)_T} \bigr],\qquad t\in[0,T],
\]
is well defined. Recall that $S$ and $\gamma\cdot S$ are
$\mathbb{Q}$-martingales. From the right-hand side of \eqref{eq:17}
and  Jensen's inequality we deduce that
\begin{eqnarray*}
-\widetilde{V}_t(x) &\leq &H \bigl(\llvert S_t-x\rrvert
\bigr)e^{(\gamma
\cdot S)_t} \leq C(\epsilon) e^{(1+\epsilon)\llvert  S_t-x\rrvert  + (\gamma
\cdot S)_t}
\\
&\leq& C(\epsilon) \mathbb{E}^{\mathbb{Q}}_t \bigl[e^{(1+\epsilon)\llvert \Psi-x\rrvert  +
(\gamma\cdot S)_T}
\bigr] \leq C(\epsilon) N_t e^{(1+\epsilon)\llvert
x\rrvert }
\end{eqnarray*}
and the global supermartingale property of $\widetilde{V}(x)$ under
$\mathbb{Q}$ follows.

We thus have shown that $V(x)=F(\llvert  S-x\rrvert )e^{-R}$ is a
supermartingale. As $F\leq1$ and $R_T=0$ we then obtain
\[
e^{-R_t} \geq F \bigl(\llvert S_t-x\rrvert \bigr)
e^{-R_t} \geq \mathbb{E}_t \bigl[F \bigl(\llvert \Psi-x
\rrvert \bigr) \bigr],\qquad x\in\mathbf{R}^n.
\]
Of course, we can replace $x$ in the inequality above with any
$\mathcal{F}_t$-measurable random variable and, in particular, with
$\mathbb{E}_t[\Psi]$. As $H$ is convex, $H(0)=0$, and
$\Vert\Psi\Vert_H<1$ we then deduce that
\begin{eqnarray*}
e^{-R_t} & \geq&\mathbb{E}_t \bigl[F \bigl( \bigl\llvert
\Psi- \mathbb{E}_t[\Psi ] \bigr\rrvert \bigr) \bigr]
\\
&= &1 - \mathbb{E}_t \bigl[H \bigl( \bigl\llvert \Psi-
\mathbb{E}_t[\Psi] \bigr\rrvert \bigr) \bigr]
\\
&=& 1- \mathbb{E}_t \biggl[H \biggl(\Vert\Psi\Vert_H
\frac{\llvert \Psi-\mathbb{E}_t[\Psi]\rrvert }{\Vert\Psi
\Vert_H} \biggr) \biggr]
\\
&\geq&1- \Vert\Psi\Vert_H \mathbb{E}_t \biggl[H \biggl(
\frac{\llvert \Psi -\mathbb{E}_t[\Psi]\rrvert }{\Vert\Psi\Vert
_H} \biggr) \biggr]
\end{eqnarray*}
and the inequality \eqref{eq:20} follows from \eqref{eq:18}.
\end{pf}

Recall that if $L$ is a $\bmo$-martingale, then the stochastic
exponential $\mathcal{E}(L)$ is a martingale, and hence, is the
density process of some probability measure $\mathbb{Q}$. Moreover, if
$\Vert L\Vert_{\bmo} \leq b$ then there is a constant $K=K(n,b)$ such
that if
$M\in\bmo(\mathbf{R}^n)$ then its Girsanov's transform $N \set M -
 \langle M,L  \rangle$ belongs to $\bmo(\mathbb{Q})$ and
\[
\frac{1}{K} \Vert N\Vert_{\bmo(\mathbb{Q})} \leq \Vert M\Vert_{\bmo}
\leq{K} \Vert N\Vert_{\bmo(\mathbb{Q})};
\]
see Theorem~3.3 in Kazamaki \cite{Kazam:94}. If $M = \beta\cdot B$,
then the
above inequality can be equivalently written as
%
\begin{equation}
\label{eq:21} \frac{1}{K} \Vert\beta\Vert_{\bmo(\mathbb{Q})} \leq \Vert\beta
\Vert_{\bmo} \leq{K} \Vert\beta\Vert_{\bmo(\mathbb{Q})}.
\end{equation}
We need a similar inequality for the $\bmo$-norm \eqref{eq:1}
associated with random variables.

\begin{Lemma}
\label{lem:3}
Let $L$ be a $\bmo$-martingale with $\Vert L\Vert_{\bmo} \leq b$,
$\mathbb{Q}$ be the probability measure with the density process
$Z=\mathcal{E}(L)$, and $\xi$ be an integrable $n$-dimensional
random variable such that $\mathbb{E}[\xi] = 0$ and
$\Vert\xi\Vert_{\bmo} <\infty$. Then $\xi$ is integrable under
$\mathbb{Q}$
and there is a constant $K=K(n,b)$ such that
%
\begin{equation}
\label{eq:22} \frac{1}{K} \bigl\Vert\xi- \mathbb{E}_{\mathbb{Q}}[\xi]
\bigr\Vert_{\bmo
(\mathbb
{Q})} \leq \Vert\xi\Vert_{\bmo} \leq{K} \bigl\Vert\xi-
\mathbb{E}_{\mathbb
{Q}}[\xi]\bigr\Vert_{\bmo(\mathbb{Q})}.
\end{equation}
\end{Lemma}

\begin{pf}
It is sufficient to prove only the first inequality
in \eqref{eq:22}. Recall that by the reverse H\"older inequality
there are constants $p_0=p_0(b)>1$ and $C_1=C_1(p_0,b)>0$ such that
\[
\bigl(\mathbb{E}_\tau \bigl[Z_T^{p_0} \bigr]
\bigr)^{1/p_0}\leq C_1 Z_\tau,
\]
for every stopping time $\tau$; see Theorem~3.1 in Kazamaki \cite{Kazam:94}.
For a random variable $\eta\geq0$, this yields
\[
\mathbb{E}^{\mathbb{Q}}_\tau[\eta] = \frac{1}{Z_\tau}
\mathbb{E}_\tau [Z_T\eta ]\leq \frac{1}{Z_\tau} \bigl(
\mathbb{E}_\tau \bigl[Z_T^{p_0} \bigr]
\bigr)^{1/p_0} \bigl(\mathbb{E}_\tau \bigl[\eta^{q_0}
\bigr] \bigr)^{1/q_0} \leq C_1 \bigl(\mathbb{E}_\tau
\bigl[\eta^{q_0} \bigr] \bigr)^{1/q_0},
\]
where $q_0 = \frac{p_0}{p_0-1} >1$.

Since $\Vert\xi\Vert_{\bmo} < \infty$, the estimate above implies that
$\xi$ is integrable under $\mathbb{Q}$ and
\[
\mathbb{E}^{\mathbb{Q}}_\tau \bigl[ \bigl\llvert \xi- \mathbb
{E}^{\mathbb{Q}}_\tau[\xi] \bigr\rrvert \bigr] \leq2
\mathbb{E}^{\mathbb{Q}}_\tau \bigl[ \bigl\llvert \xi- \mathbb
{E}_\tau[\xi] \bigr\rrvert \bigr] \leq2C_1 \bigl(
\mathbb{E}_\tau \bigl[ \bigl\llvert \xi- \mathbb{E}_\tau[\xi
] \bigr\rrvert ^{q_0} \bigr] \bigr)^{1/q_0}.
\]
This readily yields the result after we recall that for every $p\geq
1$ there is a constant $C_2 = C_2(p,n)$ such that
\[
\frac{1}{C_2}\Vert\zeta\Vert_{\bmo} \leq\Vert\zeta
\Vert_{\bmo
_p} \set \sup_\tau\bigl\Vert \bigl(
\mathbb{E}_\tau \bigl[\bigl|\zeta- \mathbb{E}_\tau [
\zeta]\bigr|^p \bigr] \bigr)^{1/p}\bigr\Vert_{\infty}\leq
C_2 \Vert\zeta\Vert_{\bmo} ,
\]
for every $n$-dimensional random variable $\zeta$ with
$\mathbb{E}[\zeta] = 0$.
\end{pf}

\begin{Lemma}
\label{lem:4}
Let $\gamma\in\mathcal{H}_\infty(\mathbf{R}^n)$ and suppose that
conditions \textup{(A1)} and \eqref{eq:16} hold and that
$\mathbb{E}[\Psi]=0$ and
\[
\Vert\Psi\Vert_{H} \leq b < 1.
\]
Then $\gamma$ is a viable demand accompanied by stocks' prices $S$
and the certainty equivalent value $R$ if and only if there exist
$\theta\in\Hbmo(\mathbf{R}^{n\times d})$ and $\eta\in
\Hbmo(\mathbf{R}^d)$ such that $(S,R,\eta,\theta)$ is a solution of
the BSDE \eqref{eq:8}--\eqref{eq:9}. Moreover, there is a constant
$K = K(n,b)>0$ such that
%
\begin{equation}
\label{eq:23} \Vert\eta\Vert_{\bmo} + \Vert\theta\Vert_{\bmo}
\leq K \Vert\Psi\Vert_{\bmo} .
\end{equation}
\end{Lemma}

\begin{pf}
Let $\gamma$ be a viable demand accompanied by stocks' prices $S$
and the certainty equivalent value $R$ and let $\eta$ and $\theta$
be as in Theorem~\ref{th:1}. Recall that $a=1$ and observe
that \eqref{eq:8} can be written as
%
\begin{eqnarray}
\label{eq:24} R_t & = &\frac{1}2 \int_t^T
\bigl( \bigl\llvert \theta^*_s\gamma_s \bigr\rrvert
^2 - \llvert \eta_s\rrvert ^2 \bigr) \,ds -
\int_t^T \eta \,dB,
\nonumber
\\[-8pt]
\\[-8pt]
\nonumber
& =& \frac{1}2 \int_t^T \llvert
\alpha_s\rrvert ^2 \,ds - \int_t^T
\eta \,dW,
\end{eqnarray}
where $\alpha= \eta+ \theta^* \gamma$ is the market price of risk
and $W = B + \int\alpha \,dt$ is a Brownian motion under the pricing
measure $\mathbb{Q}$. As $R$ is nonnegative, [see \eqref{eq:7}], and
by Lemma~\ref{lem:2},
\[
R \leq c(b) \set- \log(1 - b) >0,
\]
we deduce from the second equality in \eqref{eq:24} that
\[
\Vert\alpha\Vert^2_{\bmo(\mathbb{Q})}\leq2c(b).
\]

As the stochastic exponential $\mathcal{E}(\alpha\cdot W)$ is the
density of $\mathbb{P}$ with respect to $\mathbb{Q}$ we deduce from
Lemma~\ref{lem:3} that $\Psi$ is $\mathbb{Q}$-integrable and that
there is a constant $C_1 = C_1(n,b)$ such that
\[
\bigl\Vert\Psi- \mathbb{E}_{\mathbb{Q}}[\Psi]\bigr\Vert_{\bmo(\mathbb{Q})} \leq
C_1 \Vert\Psi\Vert_{\bmo} .
\]
As $S = S_0 + \theta\cdot W$, we have
\[
\Vert\theta\Vert_{\bmo(\mathbb{Q})} = \Vert S-S_0\Vert_{\bmo(\mathbb{Q})} =
\bigl\Vert\Psi- \mathbb{E}_{\mathbb{Q}}[\Psi]\bigr\Vert_{\bmo(\mathbb{Q})}.
\]
Then, by \eqref{eq:21}, there is a constant $C_2 = C_2(n,b)$ such
that
\[
\Vert\theta\Vert_{\bmo} \leq C_2 \Vert\theta
\Vert_{\bmo
(\mathbb{Q})} \leq C_1 C_2 \Vert\Psi
\Vert_{\bmo} .
\]
Finally, since $\theta\in\Hbmo$ and $R\geq0$, from the first
equality in \eqref{eq:24} we deduce that $\eta\in\Hbmo$ and, as
$\Vert\gamma\Vert_\infty\leq1$, that
\[
\Vert\eta\Vert_{\bmo} \leq\bigl\Vert\theta^*\gamma\bigr\Vert_{\bmo} \leq
\Vert\theta\Vert_{\bmo} .
\]
This yields \eqref{eq:23} with $K = 2 C_1C_2$.

Conversely, let $(S,R,\eta,\theta)$ be a solution of the
BSDE \eqref{eq:8}--\eqref{eq:9} with $\theta\in
\Hbmo(\mathbf{R}^{n\times d})$ and $\eta\in
\Hbmo(\mathbf{R}^d)$. In view of Theorem~\ref{th:1}, we only have to
verify the uniform integrability of the local martingales $Z =
\mathcal{E}((\eta+ \theta^*\gamma)\cdot B)$, $ZS$, and $Z
(\gamma\cdot S)$. This readily follows from $\theta$ and $\eta$
being in $\Hbmo$.
\end{pf}

\begin{pf*}{Proof of Theorem~\ref{th:2}}
In view of the homogeneity relations \eqref{eq:14}
and~\eqref{eq:15}, it is sufficient to prove the result under the
extra condition \eqref{eq:16}. Without loss of generality, we can
also assume that $\mathbb{E}[\Psi]=0$.

By Theorem~\ref{th:3} in the \hyperref[sec:bsde-with-quadratic]{Appendix},
there is a constant $b=b(n)>0$ such that if
\[
\Vert\Psi\Vert_{\bmo} \leq b,
\]
then among $(\eta,\theta)\in\bmo(\mathbf{R}^d
\times\mathbf{R}^{n\times d})$ with
%
\begin{equation}
\label{eq:25}\bigl \Vert(\eta,\theta)\bigr\Vert_{\bmo} \leq2 b,
\end{equation}
there is only one solution $(S,R,\eta, \theta)$
of \eqref{eq:8}--\eqref{eq:9} and this solution satisfies
%
\begin{equation}
\label{eq:26} \bigl\Vert(\eta,\theta)\bigr\Vert_{\bmo} \leq2 \Vert\Psi
\Vert_{\bmo} .
\end{equation}
Lemma~\ref{lem:4} then implies that $\gamma$ is a viable demand
accompanied by stocks' prices~$S$.

From Lemmas \ref{lem:2} and \ref{lem:4} and accounting
for \eqref{eq:19}, we deduce the existence of a constant $c = c(n,b)
\leq b$ such that if
\[
\Vert\Psi\Vert_{\bmo} \leq c,
\]
then every solution $(S,R,\eta, \theta)$
of \eqref{eq:8}--\eqref{eq:9} satisfies \eqref{eq:25}. Hence, there
is only one such solution, and thus stocks' prices $S$ are defined
uniquely.

Finally, from \eqref{eq:26} and \eqref{eq:10}--\eqref{eq:11} we
obtain
\begin{eqnarray*}
\Vert\sigma\Vert_{\bmo} & =& \Vert\theta\Vert_{\bmo} \leq2 \Vert
\Psi\Vert_{\bmo} ,
\\
\Vert\alpha\Vert_{\bmo} & =& \bigl\Vert\eta+ \theta^*\gamma\bigr\Vert
_{\bmo} \leq \Vert\eta\Vert_{\bmo} + \Vert\theta
\Vert_{\bmo} \leq4 \Vert \Psi\Vert_{\bmo} ,
\end{eqnarray*}
which, under \eqref{eq:16}, is precisely \eqref{eq:13}.
\end{pf*}

\subsection{Proof of Proposition \texorpdfstring{\protect\ref{prop:1}}{4.3}}
\label{sec:proof-prop-1}

The proof is divided into lemmas. We begin with a ``backward
localization'' result which does not require either (A1)
or (A2).

\begin{Lemma}
\label{lem:5}
Let $\Psi$ be a bounded $n$-dimensional random variable representing
the stocks' dividends and $\gamma$ be a viable demand for $\Psi$
accompanied by stock's prices $S$. Let $\tau$ be a stopping time
taking values in $[0,T]$. Then the predictable process
\[
\gamma'_t \set\gamma_t 1_{ \lbrace{t>\tau} \rbrace},\qquad
t\in[0,T],
\]
is a viable demand for the stocks' dividends
\[
\Psi' = \Psi1_{ \lbrace{\tau< T} \rbrace}
\]
and there are stocks' prices $S'$ for $\Psi'$ and $\gamma'$ such
that
%
\begin{equation}
\label{eq:27} S'_t = S_t,\qquad t>\tau.
\end{equation}
\end{Lemma}

\begin{pf}
To simplify notation, take the risk-aversion $a=1$. Let $\mathbb{Q}$
be the pricing measure for $\gamma$ and $S$, that is,
\[
\frac{d\mathbb{Q}}{d\mathbb{P}} = \const e^{-\int_0^T \gamma \,dS}.
\]
From the martingale property of $\gamma\cdot S$ and Jensen's
inequality, we deduce
\[
\mathbb{E}^{\mathbb{Q}} \bigl[e^{\int_0^\tau\gamma \,dS} \bigr] \leq
\mathbb{E}^{\mathbb{Q}} \bigl[e^{\int_0^T \gamma \,dS} \bigr] < \infty.
\]
This allows us to define the probability measure $\mathbb{Q}'$ such
that
\[
\frac{d\mathbb{Q}'}{d\mathbb{Q}} = \frac{e^{\int_0^\tau
\gamma\, dS}}{\mathbb{E}^{\mathbb{Q}}[e^{\int_0^\tau
\gamma \,dS}]}.
\]
Then
\[
\frac{d\mathbb{Q}'}{d\mathbb{P}} = \frac{e^{-\int_\tau^T
\gamma \,dS}}{\mathbb{E}[e^{-\int_\tau^T
\gamma \,dS}]} = \frac{e^{-\int_0^T \gamma'
\,dS}}{\mathbb{E}[e^{-\int_0^T
\gamma' \,dS}]}.
\]
Define the bounded $\mathbb{Q}'$-martingale
\[
S'_t \set\mathbb{E}^{\mathbb{Q}'} \bigl[
\Psi'|\mathcal{F}_t \bigr] = \mathbb{E}^{\mathbb{Q}'}[
\Psi1_{ \lbrace{\tau<T}
\rbrace} |\mathcal{F}_t],\qquad t\in[0,T].
\]
To show that $S'$ is a desired price process for $\Psi'$ and
$\gamma'$, we need to verify \eqref{eq:27} and the
$\mathbb{Q}'$-martingale property of $\gamma' \cdot S'$.

Since the density of $d\mathbb{Q}'/d\mathbb{Q}$ is
$\mathcal{F}_\tau$-measurable, the conditional expectations of
$\mathbb{Q}$ and $\mathbb{Q}'$ with respect to the $\sigma$-algebras
$\mathcal{F}_{\tau\vee t}$, $t\in[0,T]$, coincide. This readily
implies \eqref{eq:27}. We also deduce that if $N$ is a
$\mathbb{Q}$-martingale then
\[
N'_t \set N_t - N_{t\wedge\tau} = \int
_0^t 1_{ \lbrace{s>\tau
} \rbrace} \,dN_s,\qquad t
\in[0,T],
\]
is a $\mathbb{Q}'$-martingale. In particular, as
\[
\int_0^t \gamma'
\,dS' = \int_0^t 1_{ \lbrace{r>\tau}
\rbrace}
\gamma_r \,dS_r, \qquad t\in[0,T],
\]
we obtain that $\gamma'\cdot S'$ is a $\mathbb{Q}'$-martingale.
\end{pf}

The following lemma contains the main idea behind the proof of
Proposition~\ref{prop:1}. In its formulation, all processes and random
variables are one-dimensional.

\begin{Lemma}
\label{lem:6}
Let $B$ be a Brownian motion, $\Psi$ be a random variable different
from a constant, and $\gamma$ be a predictable process such that
\[
\bigl\llvert \Psi(\omega) \bigr\rrvert = \bigl\llvert \gamma_t(
\omega ) \bigr\rrvert = 1, \qquad \mathbb{P}[d\omega]\times \,dt\mbox{-a.s.}
\]
Then there is no a solution $(S,R,\eta, \theta)$ of the BSDE
%
\begin{eqnarray}
\label{eq:28} R_t &= &\frac{1}2 \int_t^T
\bigl(\theta_s^2 - \eta_s^2
\bigr) \,ds - \int_t^T \eta \,dB,
\\
\label{eq:29} S_t &=& \Psi- \int_t^T
\theta_s(\eta_s+\theta_s
\gamma_s) \,ds - \int_t^T \theta \,dB,
\end{eqnarray}
with bounded $S$, nonnegative $R$, and such that
%
\begin{equation}
\label{eq:30} \sign \bigl(S_t(\omega) \bigr) = -
\gamma_t( \omega), \qquad\mathbb{P}[d\omega]\times dt\mbox{-a.s.}
\end{equation}
\end{Lemma}

\begin{pf}
Suppose, on the contrary, that $(S,R,\eta, \theta)$
solves \eqref{eq:28}--\eqref{eq:29} and that $S$ is bounded, $R$ is
nonnegative, and \eqref{eq:30} holds. As in the proof of
Lemma~\ref{lem:2}, define the function
\[
F(x) \set e^{\llvert  x\rrvert } \bigl(1 - \llvert x\rrvert \bigr),\qquad x\in \mathbf{R},
\]
and observe that it is twice continuously differentiable and solves
%
\begin{equation}
\label{eq:31} F(x) - 2{F'}(x)\sign(x) + F''(x)
= 0.
\end{equation}

From  It\^o's formula and
equations \eqref{eq:28}--\eqref{eq:29} for $R$ and $S$, we deduce
that
\begin{eqnarray*}
d e^{-R_t} &=& e^{-R_t} \bigl(-\eta_t \,dB +
\tfrac{1}2 \theta^2_t \,dt \bigr),
\\
\,dF(S_t) &=& F'(S_t) \theta_t \,dB
+ \bigl(F'(S_t) \theta_t(
\eta_t + \theta_t \gamma_t) +
\tfrac{1}2 F''(S_t)
\theta^2_t \bigr) \,dt.
\end{eqnarray*}
Applying It\^o's formula to
\[
V_t = F(S_t) e^{-R_t},\qquad t\in[0,T],
\]
we then obtain that
\[
V_t = M_t + \int_0^t
e^{-R_s} A_s \,ds,
\]
where $M$ is a local martingale and
\[
A_t = \tfrac{1}2 \theta_t^2
\bigl(F(S_t) + 2F'(S_t)\gamma_t
+ F''(S_t) \bigr) =0,
\]
because of \eqref{eq:30} and \eqref{eq:31}.

Thus, $V$ is a local martingale. As $S$ is bounded and $R$ is
nonnegative, $V$ is bounded, and hence, is a martingale. Since
\[
V_T = F(S_T)e^{-R_T} = F(\Psi) = 0,
\]
we deduce that $V= 0$, and hence, that $\llvert  S\rrvert  =1$.
However, as $S$
is a continuous one-dimensional process, $S$ equals to a constant,
which contradicts the assumption that $\Psi=S_T$ is not a constant.
\end{pf}

\begin{pf*}{Proof of Proposition~\ref{prop:1}}
In view of the
self-similarity relations \eqref{eq:14}, it is sufficient to
consider the case $a=1$. Take
%
\begin{equation}
\label{eq:32} \Psi\set\sign(B_T),\qquad \gamma\set-\sign(B)
\end{equation}
and assume that $\gamma$ is accompanied by a price process $S$.
Lemma~\ref{lem:6} yields the contradiction
if
%
\begin{equation}
\label{eq:33} \sign(S_r) = \sign(B_r),\qquad r\in(0,T).
\end{equation}

Fix $r\in(0,T)$, define the stopping time
\[
\tau= \tau(r) \set\inf \lbrace{t\geq r}\dvtx B_t=0 \rbrace \wedge T,
\]
and observe that \eqref{eq:33} holds if
%
\begin{equation}
\label{eq:34} S_\tau=0\qquad \mbox{on the set } \lbrace{\tau< T} \rbrace.
\end{equation}
Indeed, in this case,
\[
S_\tau= \Psi1_{ \lbrace{\tau=T} \rbrace} = \sign(B_T)
1_{ \lbrace{\tau=T} \rbrace} = \sign(B_r) 1_{ \lbrace{\tau=T} \rbrace}
\]
and, as $S$ is a martingale under the pricing measure $\mathbb{Q}$,
we obtain
\[
S_r = \mathbb{E}^{\mathbb{Q}}[S_\tau|
\mathcal{F}_r] = \sign(B_r) \mathbb{Q}[\tau= T|
\mathcal{F}_r].
\]
This readily implies \eqref{eq:33} after we observe that, because
$r<T$ and $\mathbb{Q}$ is equivalent to $\mathbb{P}$, the
conditional probability
\[
\mathbb{Q}[\tau= T|\mathcal{F}_r] = \mathbb{Q} \Bigl[\inf
_{t\in[r,T]} \llvert B_t\rrvert >0|\mathcal{F}_r
\Bigr]
\]
is strictly positive.

In view of (A2), the stock price $S$ admits the
representation
\[
S_t(\omega) = X_t \bigl(B(\omega) \bigr) =
X_t \bigl( \bigl(B_s(\omega) \bigr)_{0\leq s\leq t}
\bigr),
\]
in terms of a continuous adapted process $X$ defined on the
canonical Wiener space of continuous functions on $[0,T]$. Define a
Brownian motion
\[
\widetilde B_t \set\int_0^t
\sign(\tau- r)\,dB_r = B_t 1_{ \lbrace{t\leq\tau} \rbrace} -
B_t 1_{ \lbrace
{t>\tau} \rbrace},\qquad t\in[0,T],
\]
and observe that, as $S$ corresponds to $\Psi$ and $\gamma$
from \eqref{eq:32}, the continuous semi-martingale
\[
\widetilde S_t \set-X_t(\widetilde B),\qquad t\in[0,T],
\]
accompanies $\widetilde\Psi$ and $\widetilde\gamma$ given by
\[
\widetilde\Psi\set-\sign(\widetilde B_T),\qquad \widetilde \gamma\set
\sign(\widetilde B).
\]
By construction,
%
\begin{equation}
\label{eq:35} S_t = X_t \bigl((B_s)_{s\leq t}
\bigr) = -\widetilde S_t,\qquad t\leq\tau
\end{equation}
and
\begin{eqnarray*}
\Psi' &\set&\sign(B_T) 1_{ \lbrace{\tau<T} \rbrace} =
\Psi1_{ \lbrace{\tau<T} \rbrace} = \widetilde\Psi1_{ \lbrace{\tau<T} \rbrace},
\\
\gamma'_t &\set&-\sign(B_t)
1_{ \lbrace{t>\tau} \rbrace
} = \gamma_t 1_{ \lbrace{t > \tau} \rbrace} = \widetilde{
\gamma}_t 1_{ \lbrace{t > \tau} \rbrace},\qquad t\in[0,T].
\end{eqnarray*}
If $\Psi'$ and $\gamma'$ are accompanied by the \emph{unique} price
process $S'$ then, by Lemma~\ref{lem:6},
\[
S'_t = S_t = \widetilde S_t,\qquad
t>\tau,
\]
and, in particular,
\[
S_\tau= \widetilde S_\tau,\qquad \tau< T,
\]
which jointly with \eqref{eq:35} implies \eqref{eq:34}. Thus, we
have a contradiction.
\end{pf*}

\begin{appendix}
\section*{Appendix: BSDE with quadratic growth in $\bmo$}
\label{sec:bsde-with-quadratic}

As before, we work on a complete filtered probability space
$(\Omega,\mathcal{F},\break (\mathcal{F}_t)_{t\in[0,T]},   \mathbb{P})$ where
$T$ is a finite time horizon and assume that (A1) holds.

Consider the $n$-dimensional BSDE:
%
\setcounter{equation}{0}
\begin{equation}
\label{eq:36} Y_t=\Xi+ \int_t^Tf(s,
\zeta_s) \,ds-\int_t^T
\zeta_s \,dB_s, \qquad t\in [0,T].
\end{equation}
Here, $Y$ is an $n$-dimensional semi-martingale, $\zeta$ is a
predictable process with values in the space of $n\times d$ matrices,
and the terminal condition $\Xi$ and the driver $f=f(t,z)$ satisfy the
following assumptions:
\begin{longlist}[(A3)]
\item[(A3)]
$\Xi$ is an integrable random variable with
values in $\mathbf{R}^n$ such that the martingale
\[
L_t \set\mathbb{E}_t[\Xi] - \mathbb{E}[\Xi],\qquad t\in[0,T],
\]
belongs to $\bmo$.
\item[(A4)]
$t\mapsto f(t,z)$ is a predictable process with
values in $\mathbf{R}^n$,
\[
f(t,0)=0,
\]
and there is a constant $\Theta>0$ such that
\[
\bigl|f(t,u)-f(t,v)\bigr| \leq \Theta\bigl(|u-v|\bigr) \bigl(|u|+|v|\bigr),
\]
for all $t\in[0,T]$ and $u,v\in\mathbf{R}^{n\times m}$.
\end{longlist}
Note that $f=f(t,z)$ has a quadratic growth in $z$.

Recall that there is a constant $\kappa= \kappa(n)$ such that, for
every martingale $M\in\bmo(\mathbf{R}^n)$,
%
\begin{equation}
\label{eq:37} \frac{1}{\kappa} \Vert M\Vert_{\bmo} \leq\Vert M
\Vert_{\bmo_1} \set \sup_{\tau}\bigl\Vert\mathbb{E}_\tau
\bigl[\llvert M_T-M_\tau\rrvert \bigr]\bigr\Vert_{\infty}
\leq\Vert M\Vert_{\bmo} ;
\end{equation}
see \cite{Kazam:94}, Corollary~2.1, page 28. Hereafter, we fix the
constants $\kappa$ and $\Theta$ from \eqref{eq:37} and (A4)
and use the $\bmo$-martingale $L$ from (A3).

\begin{Theoremm}
\label{th:3}
Assume \textup{(A1)}, \textup{(A3)} and \textup{(A4)}. If
%
\begin{equation}
\label{eq:38} \Vert L\Vert_{\bmo} < \frac{1}{8\kappa\Theta},
\end{equation}
then there is $\zeta\in\Hbmo$ solving \eqref{eq:36} and such that
%
\begin{equation}
\label{eq:39} \Vert\zeta\Vert_{\bmo} \leq2 \Vert L\Vert_{\bmo}
.
\end{equation}
Moreover, if \eqref{eq:38} holds and $\zeta' \in\Hbmo$ is another
solution to \eqref{eq:36} such that
%
\begin{equation}
\label{eq:42} \bigl\Vert\zeta'\bigr\Vert_{\bmo} \leq
\frac{1}{4\kappa\Theta},
\end{equation}
then $\zeta= \zeta'$.
\end{Theoremm}

\begin{Remarkk}
\label{rem:5}
Theorem~\ref{th:3} extends Proposition~1 in Tevzadze \cite{Tevz:08}, where
the terminal condition $\Psi$ is supposed to have sufficiently small
$\mathbf{L}_{\infty}$-norm. A similar extension to the case
$\Psi\in\bmo$ has been obtained independently in Proposition~2.1 of
Frei \cite{Frei:14}, however, with slightly different constants.

We are unaware of any general result on the \emph{global} uniqueness
of a \emph{local} solution $\zeta$ from Theorem~\ref{th:3}; that is,
the uniqueness of $\zeta$ in the whole space $\Hbmo$, without the
constraint \eqref{eq:42}. This highlights the relevance of
Theorem~\ref{th:2} which, to the best of our knowledge, is the first
example of a coupled system of quadratic BSDEs where such uniqueness
is established.
\end{Remarkk}

We divide the proof of Theorem~\ref{th:3} into lemmas.

\begin{Lemmaa}
\label{lem:7}
Assume \textup{(A1)}, \textup{(A3)} and \textup{(A4)}. For $\zeta\in
\Hbmo$, there is unique $\zeta'\in\Hbmo$ such that
%
\begin{equation}
\label{eq:40} \bigl(\zeta'\cdot B \bigr)_t =
\mathbb{E}_t \biggl[\Xi+ \int_0^Tf(s,
\zeta _s)\,ds \biggr] - \mathbb{E} \biggl[\Xi+ \int
_0^Tf(s,\zeta_s)\,ds \biggr].
\end{equation}
Moreover,
%
\begin{equation}
\label{eq:41} \bigl\Vert\zeta'\bigr\Vert_{\bmo} \leq\Vert L
\Vert_{\bmo} + 2\kappa\Theta \Vert\zeta\Vert_{\bmo}
^2.
\end{equation}
\end{Lemmaa}

\begin{pf}
Define the martingale
\[
M_t \set\mathbb{E}_t \biggl[\int_0^Tf(s,
\zeta_s)\,ds \biggr] - \mathbb{E} \biggl[\int_0^Tf(s,
\zeta_s)\,ds \biggr].
\]
For a stopping time $\tau$, we deduce from (A4) and
It\^o's isometry that
\begin{eqnarray*}
\mathbb{E}_{\tau} \bigl[\llvert M_T - M_\tau
\rrvert \bigr] & =& \mathbb{E}_\tau \biggl[ \biggl\llvert \int
_\tau^T f(s,\zeta_s)\,ds -
\mathbb{E}_\tau \biggl[\int_\tau^T
f(s, \zeta_s) \,ds \biggr] \biggr\rrvert \biggr]
\\
& \leq&2\mathbb{E}_{\tau} \biggl[\int_{\tau}^T
\bigl\llvert f(s,\zeta _s) \bigr\rrvert \,ds \biggr] \leq 2\Theta
\mathbb{E}_{\tau} \biggl[\int_{\tau}^T
\llvert \zeta _s\rrvert ^2 \,ds \biggr]
\\
&=& 2\Theta\mathbb{E}_{\tau} \biggl[ \biggl\llvert \int
_\tau^T \zeta \,dB \biggr\rrvert ^2
\biggr].
\end{eqnarray*}
Accounting for \eqref{eq:37}, we obtain
\[
\Vert M\Vert_{\bmo} \leq2\kappa\Theta\bigl(\Vert\zeta\cdot B\Vert
_{\bmo} \bigr)^2= 2\kappa\Theta\Vert\zeta\Vert_{\bmo}
^2.
\]
This shows that the martingale on the right-hand side
of \eqref{eq:40} belongs to $\bmo$. In view of (A1) it then
admits an integral representation as $\zeta'\cdot B$ for some
$\zeta'\in\Hbmo$. We clearly have that $\zeta'$ is unique in
$\Hbmo$ and
\[
\bigl\Vert\zeta'\bigr\Vert_{\bmo} = \bigl\Vert\zeta'\cdot B
\bigr\Vert_{\bmo} \leq \Vert L\Vert_{\bmo} + \Vert M
\Vert_{\bmo} .
\]
\upqed\end{pf}

Lemma~\ref{lem:7} allows us to define the map
\[
F\dvtx \Hbmo \rightarrow\Hbmo
\]
such that $\zeta'=F(\zeta)$ is given by \eqref{eq:40}.

\begin{Lemmaa}
\label{lem:8}
Assume \textup{(A1)}, \textup{(A3)} and \textup{(A4)}. Let $\zeta$ and
$\zeta'$ be in $\Hbmo$. Then
\[
\bigl\Vert F(\zeta)-F \bigl(\zeta' \bigr)\bigr\Vert_{\bmo} \leq2
\kappa \Theta\bigl\Vert\zeta-\zeta'\bigr\Vert_{\bmo} \bigl(\Vert\zeta
\Vert_{\bmo} + \bigl\Vert\zeta'\bigr\Vert_{\bmo} \bigr).
\]
\end{Lemmaa}

\begin{pf} We have
\[
\bigl\Vert F(\zeta)-F \bigl(\zeta' \bigr)\bigr\Vert_{\bmo} = \Vert M
\Vert_{\bmo} ,
\]
where
\[
M_t \set\mathbb{E}_t \biggl[\int_0^T
\bigl(f(s,\zeta_s)-f \bigl(s,\zeta '_s
\bigr) \bigr)\,ds \biggr] - \mathbb{E} \biggl[\int_0^T
\bigl(f(s,\zeta_s)-f \bigl(s,\zeta'_s
\bigr) \bigr)\,ds \biggr].
\]
For a stopping time $\tau$, we deduce from (A4) that
\[
\mathbb{E}_{\tau} \bigl[\llvert M_T - M_\tau
\rrvert \bigr] \leq 2\Theta\mathbb{E}_{\tau} \biggl[\int
_{\tau}^T \bigl\llvert \zeta_s -
\zeta'_s \bigr\rrvert \bigl(\llvert
\zeta_s \rrvert + \bigl\llvert \zeta'_s
\bigr\rrvert \bigr) \,ds \biggr].
\]
Cauchy's inequality and  It\^o's isometry then yield
\[
\mathbb{E}_{\tau} \bigl[\llvert M_T - M_\tau
\rrvert \bigr] \leq 2\Theta \bigl\Vert\zeta-\zeta'\bigr\Vert_{\bmo}
\bigl(\Vert\zeta\Vert_{\bmo} + \bigl\Vert \zeta'
\bigr\Vert_{\bmo} \bigr).
\]
The result now follows from \eqref{eq:37}.
\end{pf}

\begin{pf*}{Proof of Theorem~\ref{th:3}}
From Lemma~\ref{lem:7}, we deduce that $F$ maps the ball of the
radius $R \set\frac{1}{4\kappa\Theta}$ into the ball of the radius
\[
R' = \Vert L\Vert_{\bmo} + 2\kappa\Theta R^2 <
R.
\]
From Lemma~\ref{lem:8}, we obtain that $F$ is a contraction on the
ball of the radius $R'$: if $\zeta,\zeta'\in\Hbmo$ and
$\max(\Vert\zeta\Vert_{\bmo} ,\Vert\zeta'\Vert_{\bmo} )\leq
R'$, then
\begin{eqnarray*}
\bigl\Vert F(\zeta)-F \bigl(\zeta' \bigr)\bigr\Vert_{\bmo} &\leq&2
\kappa\Theta \bigl\Vert\zeta-\zeta'\bigr\Vert_{\bmo} \bigl(\Vert\zeta
\Vert_{\bmo} +\bigl\Vert \zeta'\bigr\Vert_{\bmo} \bigr)
\\
& \leq&\frac{R'}{R} \bigl\Vert\zeta-\zeta'\bigr\Vert_{\bmo} .
\end{eqnarray*}

Banach's fixed-point theorem now implies the existence and
uniqueness of $\zeta\in\Hbmo$ such that $\Vert\zeta\Vert_{\bmo}
\leq R$ and
$F(\zeta) = \zeta$. The estimate \eqref{eq:39} for $\zeta$ follows
from \eqref{eq:41}:
\[
\Vert\zeta\Vert_{\bmo} \leq\Vert L\Vert_{\bmo} + 2\kappa\Theta
\Vert\zeta \Vert_{\bmo} ^2 \leq \Vert L\Vert_{\bmo} +
\tfrac{1}2 \Vert\zeta\Vert_{\bmo} .
\]
It only remains to observe that the fixed points of $F$ are in
one-to-one correspondence with the solutions $\zeta$
to \eqref{eq:36} such that $\zeta\cdot B \in\bmo$.
\end{pf*}
\end{appendix}

%
%
%





\printaddresses
\end{document}